%% file: paper.tex
\documentclass[sigplan]{acmart}

\usepackage[normalem]{ulem}
\usepackage{subcaption}
\usepackage{xspace}
\usepackage{amsfonts}
\usepackage{amsmath}
\usepackage{cleveref}
\usepackage{multirow}
\usepackage{paralist}
\usepackage{soul}
\usepackage{array}

\usepackage[most]{tcolorbox}

\newtcblisting{console}{
  colback=black!5,
  colframe=black!20,
  boxrule=0.5pt,
  arc=2pt,
  left=2pt,right=2pt,top=2pt,bottom=2pt,
  fontupper=\ttfamily\small,
  breakable,
  before skip=\smallskipamount,
  after skip=\smallskipamount,
  listing only,
}

\usepackage{url}

\newcommand{\sys}{Wayfinder\xspace}
\newcommand{\algo}{\textsc{DeepTune}\xspace}

\usepackage{pgfplots}
\usepackage{tikz}
\newcommand*\circled[1]{\protect\tikz[baseline=(char.base)]{
	\node[shape=circle,gray,draw,inner sep=0.15pt,line width=0.3mm] (char)
	{\textcolor{black}{#1}};}}

\usetikzlibrary{positioning,petri}

\copyrightyear{2026}
\acmYear{2026}
\setcopyright{cc}
\setcctype{by}
\acmConference[EUROSYS '26]{21st European Conference on Computer Systems}{April 27--30, 2026}{Edinburgh, Scotland Uk}
\acmBooktitle{21st European Conference on Computer Systems (EUROSYS '26), April 27--30, 2026, Edinburgh, Scotland Uk}
\acmDOI{10.1145/3767295.3803589}
\acmISBN{979-8-4007-2212-7/2026/04}

\ccsdesc[500]{Software and its engineering~Operating systems}
\ccsdesc[500]{Software and its engineering~Software configuration management and version control systems}

\keywords{Operating Systems, Specialization}

\begin{document}

\title{\sys:\\Automated Operating System Specialization}

\author{\rm Alexander Jung}
\affiliation{%
  \institution{Lancaster University}
  \country{United Kingdom}
}
\affiliation{
\institution{Unikraft GmbH}
\country{Germany}
}

\author{\rm Cezar Crăciunoiu}
\affiliation{
\institution{Politehnica University of Bucharest}
\country{Romania}
}

\author{\rm Nikolaos Karaolidis}
\affiliation{
\institution{The University of Manchester}
\country{United Kingdom}
}

\author{\rm Hugo Lefeuvre}
\affiliation{
\institution{The University of British Columbia}
\country{Canada}
}

\author{\rm Daniel Oñoro Rubio}
\affiliation{
\institution{NEC Laboratories Europe}
\country{Germany}
}

\author{\rm Felipe Huici}
\affiliation{
\institution{Unikraft GmbH}
\country{Germany}
}

\author{\rm Charalampos Rotsos}
\affiliation{
\institution{Lancaster University}
\country{United Kingdom}
}

\author{\rm Pierre Olivier}
\affiliation{
\institution{The University of Manchester}
\country{United Kingdom}
}

\input{00-abstract}

\maketitle
\pagestyle{plain}

\input{01-introduction}
\input{02-motivation}

\input{03-design-and-implementation}
\input{04-evaluation}
\input{05-related-works}
\input{06-conclusion}

\input{07-acknowledgements}

\input{A-artifact-evaluation}

\bibliographystyle{ACM-Reference-Format}
\bibliography{bib}

\end{document}

%% file: 00-abstract.tex
\begin{abstract}
Operating system specialization is a well-known approach to optimize a specific application's performance, memory usage, security, or other important metrics.
Specializing an OS for an application is typically a manual process that requires great expertise.
Specialization through configuration lends itself well to automation; however, it is challenging due to the sheer size of the configuration space of modern OSes, the difficulty to quantify that space, the long time it takes to evaluate a configuration, and the large number of invalid configurations.
Hence, existing attempts at specializing OSes automatically are limited to switching features on and off to minimize memory consumption or attack surface, and cannot target metrics such as performance.

We present \sys, a framework specializing the configuration of OSes completely automatically and without expert knowledge.
It can specialize all aspects of an OS configuration (compile-/boot-/run-time) towards any quantifiable performance, resource consumption, or security metric, for an application processing a given workload on a given hardware setup.
\sys consists of an automated OS benchmarking platform, and a neural network-based search algorithm driving the specialization process.
This is achieved by learning on the fly which configuration parameters and values impact performance the most, and which ones lead to runtime failures.
Optionally, a model pre-trained on one application can be reused to accelerate the specialization of related applications.
We evaluate \sys on two OSes, four applications, and two target metrics: \sys fully automatically identifies specialized configurations with up to 24\% application performance improvement and 8.5\% memory usage reduction compared to default configurations.
We highlight the benefits of our neural network, reaching good solutions faster than competing approaches (random and Bayesian), and successfully transferring knowledge between related applications.
\end{abstract}

%% file: 01-introduction.tex
\section{Introduction}

The general-purpose nature of mainstream Operating Systems (OSes) comes at the expense of performance, resource consumption, or security~\cite{SINGLE_USER_OS, EXOKERNEL, ARRAKIS, IX, BARRELFISH, FLEXOS_HOTOS}.
Thus, a common technique to trade off the cost of generality is \emph{OS specialization}~\cite{NUCLEUS_OS, SINGLE_USER_OS, APP_SPECIFIC_OS, EXOKERNEL}, the process of tailoring an OS towards specific use-cases such as applications, workloads, or hardware platforms.
OS specialization is a popular systems research topic, with works targeting I/O~\cite{SANDSTORM, ARRAKIS, IX, LIBRETTOS, UNIKRAFT, UKL}, GPU or hardware accelerators~\cite{GPUNET, GPUFS, OMNIX}, resource usage~\cite{MIRAGE_ASPLOS, LIGHTVM, FIRECRACKER, UNIKRAFT, LUPINE}, security~\cite{FLEXOS_ASPLOS, GRAPHENE_SGX, XCONTAINER, ISOLSEC}, compatibility~\cite{GRAPHENE, SCONE, ELEOS, HERMITUX, HERMITUX_TC}, or extensibility~\cite{BASCULE}.

Most OS specialization efforts rely on expert developers manually tuning or re-implementing specific OS subsystems.
Unfortunately, such approaches depend on expert domain knowledge, preventing the vast majority of users from reaping the benefits of these techniques.
To address this problem, we need more automation in OS specialization.

We focus on OS specialization \emph{through configuration}, where an OS enables users to customize its behavior and tune its functionality for specific use-cases through compile-, boot-, or runtime configuration parameters.
Although the benefits of configuration specialization can be arguably more modest than those obtained by redesigning parts of an OS, they have been shown to be significant in many scenarios~\cite{LUPINE, WAYFINDER, ISOLSEC, LOHMANN, REFRACT}.
Further, this approach lends itself well to automation, and can be achieved without expert knowledge.
Although prior work has explored automatically switching features on and off to reduce resource consumption~\cite{PRACTICAL_OS_TAILORING, ML_CONF_SIZE, LOHMANN} or attack surface~\cite{ISOLSEC, ISOLSEC2, REFRACT}, none can handle the challenges of the vast and complex configuration space offered by modern OSes when targeting other metrics such as performance.

We present \sys, a framework for specializing the configuration of OSes, such as Linux, for applications \emph{fully automatically}, and \emph{without expert knowledge}.
\sys automatically specializes any aspect of an OS configuration, such as compile-time, boot-time, and runtime parameters, towards any quantifiable performance, resource consumption, security, or compatibility metric, for a target application processing a given workload on a given hardware setup.
The framework includes an automated OS configuration, build, and benchmarking platform, designed to support reproducible testing.
At its core, \sys builds on \algo, a novel neural network-based optimization algorithm, to drive the specialization process.
The key idea of \algo is to direct the search by selecting interesting candidate configurations through predicting 1) their performance on a target metric (e.g., throughput, memory usage) and 2) their likelihood to be valid (e.g., to compile, not to crash at runtime).

\begin{figure}
      \center
      \includegraphics[width=.40\textwidth]{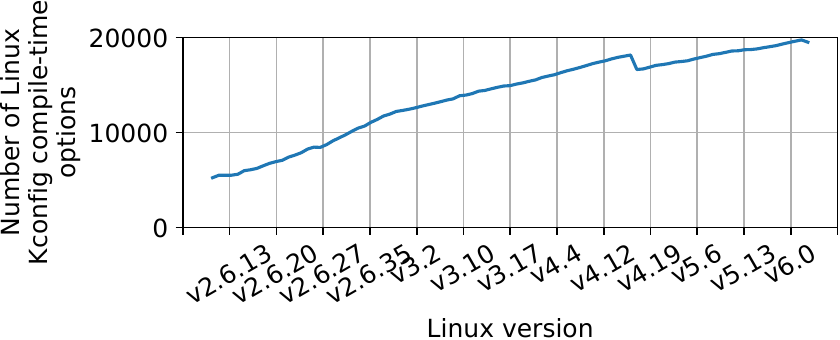}
      \caption{Linux compile-time configuration space over time.}
      \label{fig:linux-plot}
\end{figure}

\sys addresses several challenges.
First, the configuration space of modern OSes is extremely large.
As shown in Figure~\ref{fig:linux-plot}, Linux 6.0 exhibits about 20\,000 compile-time options, on top of boot-time and runtime parameters.
Some parameters take arbitrary numbers as values, which exacerbates the issue: overall, it is not possible to explore every configuration.
Second, the search space is hard to define without expert knowledge.
Many kernel parameters offer little to no documentation: parameter types and valid ranges of values are unknown.
Third, evaluating a configuration takes time, as it potentially requires building a kernel image, booting it, and running a test to evaluate the target metric.
The time required to evaluate a configuration can vary significantly, and an automated OS tuning system should optimize the overall time to discover a specialized configuration.
Finally, the search space contains many configurations that are valid on paper (e.g., on Linux, they satisfy constraints checked by KConfig), but lead to failure at compile-time, boot-time, or runtime.
Our evaluation demonstrates that, when using a naive (random search) approach to optimize the configuration of a Linux OS, approximately one third of all attempts fail, which represents a significant number of wasted search iterations.
Combined with slow evaluation, this severely limits the number of configurations that can be explored in a given time budget.

\sys's design tackles these challenges as follows.
For OSes with large and hard-to-qualify search spaces (e.g., Linux), we determine the search space offline using a heuristic algorithm that infers both types and ranges for each OS parameter.
Then, our online exploration relies on \algo, a neural network optimization algorithm that progressively learns the OS configuration parameters and values with the most significant performance impact for a target metric.
To speed up the search, \algo learns to avoid parameter values that are likely to trigger failures, a feature that competing methods, such as random search or Bayesian optimization, lack.
By default, the learning process starts from scratch for every application/metric to specialize towards.
Optionally, after training a model to optimize for a given application, transfer learning can be applied, i.e., the model can be reused to accelerate exploration on other applications with similar characteristics.

We apply \sys to specialize the configuration of two OSes, Linux and Unikraft~\cite{UNIKRAFT}, for popular cloud applications.
We optimize for two target metrics: application performance and memory usage.
Unlike competitors such as causal reasoning~\cite{UNICORN} or Bayesian optimization~\cite{WAYFINDER}, \sys scales to the vast design space of modern OSes such as Linux, and discovers specialized configurations several times faster from a random search baseline.
\sys is efficient at predicting failures, reducing crash rates from 30\% to 10-25\%.
Lastly, we demonstrate that transfer learning is effective in accelerating search time and reducing crash rates to less than 10\%.
For example, with a model trained on Redis and applied to Nginx, the search is sped up by 24\%, with crash rates below 5\%.

Overall, this paper makes two core contributions:
\begin{itemize}
      \item \sys, an evaluation platform able to configure, build, run, and benchmark OSes automatically and without expert knowledge.
      \item \algo, a novel neural network optimization algorithm that drives \sys's specialization process.
      %\item The evaluation of \sys and \algo.
\end{itemize}

%% file: 02-motivation.tex
\section{Background, Challenges, and Motivation}
\label{sec:motivation}

\subsection{Specialization through Configuration}

Specialization optimizes an OS for a given use-case (e.g., application, workload, hardware platform) and a given metric (e.g., performance, resource usage).
There are many approaches to OS specialization~\cite{ARRAKIS, OMNIX, MIRAGE_ASPLOS, FLEXOS_ASPLOS, ELEOS, BASCULE, UNIKRAFT, ISOLSEC}.
We focus on specialization \emph{through configuration}~\cite{LIGHTVM, LUPINE, UKL, HERMITUX, PRACTICAL_OS_TAILORING, ML_CONF_SIZE, ISOLSEC, ISOLSEC2, XCONTAINER}, where an OS is specialized by fine-tuning its compile-time, boot-time, and runtime parameters.

Prior work explored OS specialization through configuration with the goals of reducing resource usage (e.g., memory footprint)~\cite{LIGHTVM, LUPINE, UKL, HERMITUX, PRACTICAL_OS_TAILORING, ML_CONF_SIZE, LOHMANN} and attack surface~\cite{ISOLSEC, ISOLSEC2, XCONTAINER, REFRACT}.
Some of these approaches are manual~\cite{LUPINE, UKL, HERMITUX, UNIKRAFT, XCONTAINER}, hence they do not scale to more than a few applications given the size of the configuration space of real-world OSes such as Linux~\cite{XCONTAINER, LUPINE, UKL}.
To address this, other works explored automated methods to specialize OS configurations~\cite{PRACTICAL_OS_TAILORING, ISOLSEC, ISOLSEC2, ML_CONF_SIZE, REFRACT, LOHMANN}.
These efforts primarily focus on security (attack surface) or resource optimization (memory footprint). Their objective is thus to determine, for a given workload, which compile-time configuration parameters (e.g., kernel features or modules) can be disabled, and which ones are essential for the workload.

\begin{table}
    \center
    \caption{Configuration space for Linux 6.0, including boot time options (kernel command line parameters), runtime options (writable files in \texttt{/proc/sys} and \texttt{/sys}), and compile-time options (obtained by parsing the \texttt{Kconfig} hierarchy).}
    \includegraphics[width=.4\textwidth]{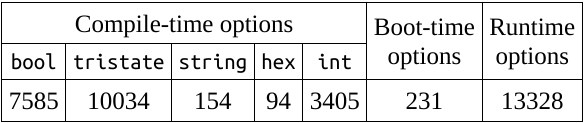}
    \label{tab:linux-parameter-types}
\end{table}

These approaches are not generic, and do not apply to performance specialization towards performance.
Beyond enabling/disabling only compile-time features, we need to consider the full set of kernel options, many of them taking arbitrary values, with unclear/undocumented ranges of validity.
To scale to many applications we cannot assume expert knowledge (i.e. no filtering of relevant options).
In short, the size of the exploration space becomes quasi-infinite.
Consider for example the size of the configuration space for the Linux kernel, illustrated in \Cref{tab:linux-parameter-types}.
The compile-time configuration of Linux includes more than 3000 options (out of the 20\,000 total) taking arbitrary integers.
Furthermore, due to its general-purpose nature, Linux must cater to a wide range of workloads, and thus many performance-critical configuration options are available only in the form of run-time parameters.
Setting these parameters correctly to maximize performance, even manually and with expert knowledge, is complex, as demonstrated by the many performance tuning guides available online~\cite{TUNING1, TUNING2, TUNING3, GUIDE_SOMAXCONN, TUNING5, TUNING6, TUNING7, TUNING8, TUNING9}.
For these reasons, automated OS specialization through configuration with performance goals has not been explored in many prior works.

\subsection{Optimizing OS Configurations for Performance}

% TODO(Pierre): Remember why v4.19 and not v4.0
To further motivate OS performance specialization through configuration, we randomly generate 800 configurations of Linux v4.19 and evaluate their performance.
For each random configuration, we configure and run a corresponding kernel in a KVM virtual machine, where we execute and benchmark an Nginx web server with \texttt{wrk}.
We run these experiments on an Intel Xeon E5-2697 v2 (2x24 cores@2.70~GHz, 128~GB RAM).
We want to obtain 800 valid configurations so when one fails (about a third of randomly generated configurations crash at runtime), we re-generate a random configuration until we obtain a valid one.

\begin{figure}
    \center
    \includegraphics[width=0.45\textwidth]{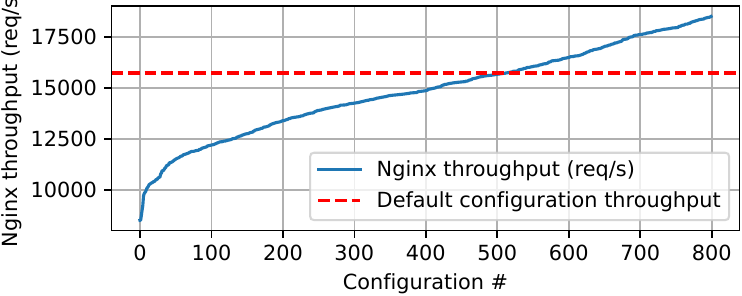}
    \caption{Nginx throughput for 800 random configurations of the Linux kernel.}
    \label{fig:linux_throughput}
\end{figure}

\Cref{fig:linux_throughput} presents the results comparing the performance of each random configuration to the default one.
The configurations are sorted in ascending performance order.
The performance varies by as much as 80\%, from under 10K up to 18K~req/s.
A key observation is that the throughput of the fastest configuration is 12\% higher than the throughput obtained with the default configuration, even though we explored only a tiny fraction of the design space.

This motivates the usefulness of optimizing OS configurations for metrics such as performance.
Still, the random search approach taken here is suboptimal: on this small space, 64\% of the configurations perform worse than the default configuration.
Further, one third of the configurations lead to situations in which the kernel cannot build/boot or crashes/hangs at runtime (referred to as a whole as \emph{crashes} in the rest of this paper), wasting resources.
Given the sheer size of the exploration space, and the lack of consideration for valid/invalid configurations, finding specialized configurations through random search would be unacceptably slow.
Hence, we propose \sys to drive the configuration space exploration using efficient optimization algorithms.

\subsection{Optimization Algorithms for Configuration-Based OS Specialization} \label{subsec:motiv-algo}

We approach the search for specialized OS configurations as an optimization problem.
Past work used various approaches, such as Bayesian optimization~\cite{WAYFINDER, COSE}, causal reasoning~\cite{UNICORN}, and random search~\cite{WAYFINDER}.
However, several issues cause these approaches to be inapplicable or inefficient:

\noindent\textbf{Scalability.} Bayesian optimization relies on Gaussian processes, which typically have a computational complexity of $O(n^3)$, and $O(n^2)$ for memory consumption. The complexity of causal analysis algorithms ranges from $O(n^4)$ to $O(n^3)$~\cite{pmlr-v161-wienobst21a}, which limits them to relatively small problems.

\noindent\textbf{Lack of incremental training.} Adding new data points and updating the model typically requires retraining a Gaussian process in the case of Bayesian optimization. It also requires recomputing the casual graph for causal inference. Combined with the scalability issue, this leads the cost of each iteration of the optimization to grow exponentially.

\noindent\textbf{Difficulty to fit both categorical and numerical parameters for high-dimensional data.} Another well-known limitation of Bayesian optimization is its poor performance on problems with categorical input features~\cite{GARRIDOMERCHAN202020} (i.e. features with a fixed set of values), and with high-dimensional inputs.

Overall, the very large configuration space we target makes both causal inference and Bayesian optimization poor fits.
Random search performs well on such large spaces, but it suffers from a high crash rate for this problem (one third of the configurations randomly sampled fail).
We further demonstrate these limitations in the evaluation (Section~\ref{sec:evaluation}).

%% file: 03-design-and-implementation.tex
\section{\sys: Design and Implementation}
\label{sec:design-and-implementation}

\begin{figure*}[ht]
% \begin{minipage}[c]{0.33\textwidth} 
%         \centering
%         \includegraphics[width=0.95\textwidth]{include/wayfinder-config-engine-tnr.pdf}
%         \caption{Overview of \sys's configuration generation API, filter API and
%                scheduler API which work in series towards generating and filtering
%                the large number of possible configuration permutations given an OS model
%                experiment.}
%         \label{fig:wayfinder_config_engine}
% \end{minipage}\hfill
\begin{minipage}[c]{0.48\textwidth}
        \centering
        \includegraphics[width=\textwidth]{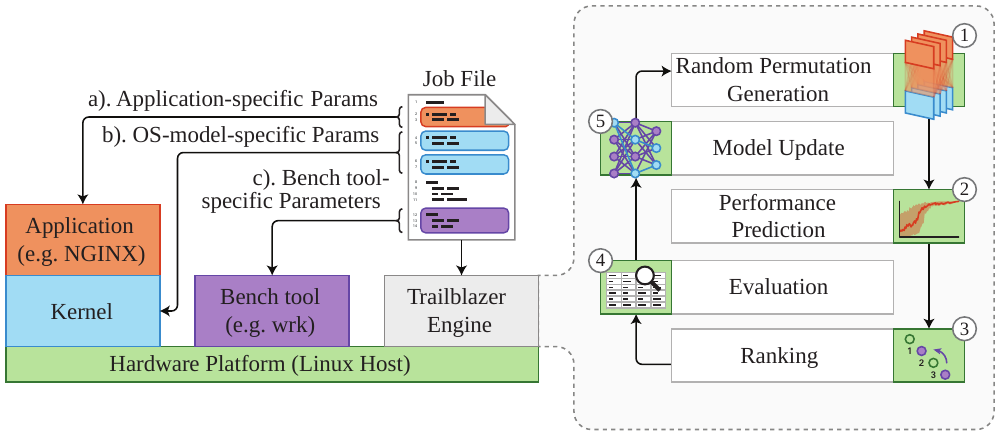}
        \caption{\algo's general process.}
        \label{fig:deeptune_dia}
        \vspace{\fill}
\end{minipage}\hfill
\begin{minipage}[c]{0.5\textwidth}
        \centering
        \includegraphics[width=\textwidth]{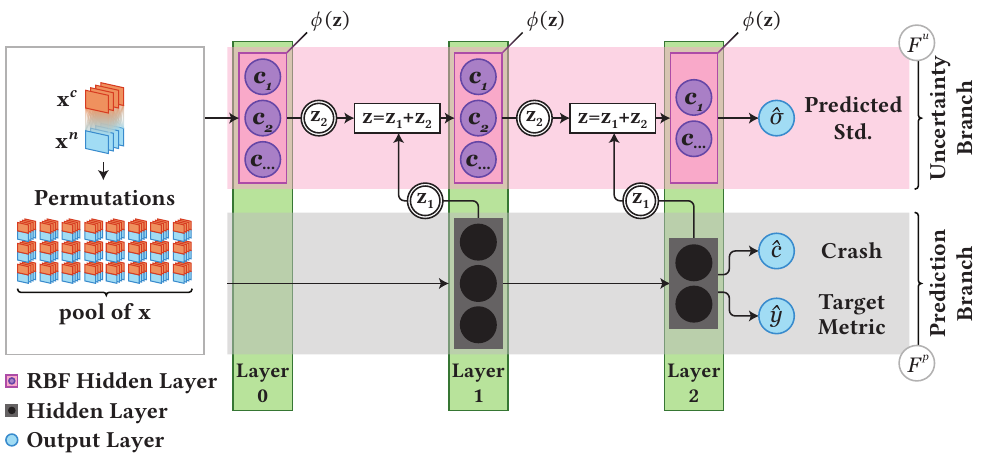}
        \caption{\algo Model overview.}
        \label{fig:deeptune_model}
\end{minipage}
\end{figure*}

The sheer size of the exploration space and the impracticability of relying on expert knowledge motivate the need for an \emph{automated} exploration platform.
That platform can be programmed to automatically configure, build, and benchmark series of OS images with the goal of specializing them towards specific applications and workloads.

\paragraph{Overview}

\sys is composed of a benchmarking pipeline (\S\ref{subsec:overview-bench-platform}) that automatically builds, configures, and benchmarks OS and application images, and of \algo (\S\ref{sec:algo}), an optimization algorithm that drives the exploration to find specialized OS configurations for a given application.

\subsection{Automated Benchmarking Pipeline} \label{subsec:overview-bench-platform}

\sys takes as input YAML files representing the configuration space of the target OS (\emph{job files}, discussed in \S\ref{sec:defining-the-space}) and scripts describing how to build and benchmark images of that OS---including the application under test.

With \sys the exploration process specializing an OS for a given application consists of iteratively executing the following core loop: 1) build and boot an OS image based on a given configuration in a VM; 2) benchmark the target application running on that OS image; and 3) determine the next configuration to consider.
The user provides the platform with a time budget or a number of iterations to run, after which the best configuration found is returned.
\sys offers a modular API to ease the integration of pluggable search algorithms, which accomplishes step 3.
These algorithms drive the OS specialization process for an application and workload/metric by deciding what configuration to explore next using various approaches, and we currently have support for the following:

\begin{compactitem}
  \item \emph{Random search}: each subsequent configuration to explore is generated randomly without considering the exploration history.
  \item \emph{Grid search}: all possible configurations are explored systematically, one parameter value after the other.
  \item \emph{Bayesian optimization}: an approach balancing exploration (trying new configurations) and exploitation (trying to optimize further configurations known to perform well) using a probabilistic model built from the exploration history.
  \item \emph{\algo}: an ML-based approach balancing exploration and exploitation described below in \S\ref{sec:algo}.
\end{compactitem}

These algorithms interact with \sys through an API exposing various information such as the history of configurations explored, the corresponding performance\footnote{In the following, we use the term ``performance'' to refer to the output of one or more user-provided metrics, which may refer to any quantifiable measure(s) such as throughput, latency, memory usage, image size, etc.} results, which configurations resulted in build failure or runtime crashes, etc.
Once a configuration is selected for evaluation, the platform creates two corresponding internal tasks: a build task to create the OS image, and a test task to measure its performance.
An optimization here is that the build task can be skipped if the differences between the current configuration to explore and the previous one only relate to runtime parameters and not boot/compile-time ones.

\sys is built in 15K LoC of Go as a collection of microservices.
It uses off-the-self components for persistence, monitoring and logging.
The platform runs on a Linux host, and benchmarked OS images execute on top of QEMU/KVM.

% The system has been designed to retrieve its state by connecting to a central Redis service that maintains a queue of all ongoing experiments, divided into build jobs, and test (benchmark) jobs.
% Should an instance of \sys crash or go offline, simply restarting or re-connecting the service allows it to regain state immediately from the Redis queue and continue its operation.
% All benchmark results are saved to a central Postgres instance.

% \sys can optionally exploit multicore CPUs to benchmark multiple OS images in parallel while minimizing co-location disturbances.
% The system automatically analyses the hardware configuration of the host to determine its topology.
% This information is fed, along with job requirements, to the job scheduler, which can make informed decisions when dispatching build and test jobs.
% When available, \sys leverages Intel's Cache Allocation Technology~\cite{INTEL_CAT} and Memory Bandwidth Allocator~\cite{INTEL_MAB} mechanisms.

\subsection{\algo}
\label{sec:algo}
\algo is the AI algorithm that drives \sys's automatic optimization.
With our key assumption that the user has no performance-tuning expertise, faced with the previously-described gigantic configuration space, the intuition behind \algo's design is that we need to combine \emph{exploration}, i.e., trying new parameters to find those that matter, and \emph{exploitation}, i.e., optimizing parameters identified as important.
This approach automatically subsets the configuration space to focus on the most impactful parameters (vs. heuristics that may require variable amounts of expertise), while ensuring that it does not become trapped in local optima.
\algo addresses the limitations we observe in existing automatic optimization algorithms such as Bayesian optimization and causal inference~\cite{UNICORN} (see \S\ref{subsec:motiv-algo}) to achieve both high accuracy and scalability.
It does so through (1) a new Neural Network (NN) model design that predicts behaviors of randomly generated permutations, and (2) a new scoring function that ranks configurations based on the predictions.
The former (1) is embodied in the \algo Model (DTM), a multitask NN that predicts runtime performance and the likelihood that a configuration will crash at runtime.
The DTM also provides a measure of uncertainty through a new mechanism based on Radial Basis Function (RBF) layers~\cite{Lee1999} and the Chamfer distance loss~\cite{fangsg17}.
The scoring function (2) ranks candidate configurations by merging the model prediction, the predicted uncertainty, and the dissimilarity in relation to known configurations.

Figure \ref{fig:deeptune_dia} illustrates the key components of \algo.
\algo starts with the random generation of a diverse pool of permutation candidates {\circled{\textbf{1}}} from \sys.
The DTM then estimates the performance of these candidates {\circled{\textbf{2}}}, and the scoring function {\circled{\textbf{3}}} ranks them.
The top permutation is evaluated by \sys~{\circled{\textbf{4}}, and the DTM is updated~{\circled{\textbf{5}}.
The algorithm iterates for a predetermined number of cycles or until a performance goal is achieved.

\noindent \textbf{The \algo Model (DTM).} This section is designed to be read along with Figure~\ref{fig:deeptune_model}. The DTM (Figure~\ref{fig:deeptune_model}) is a multitask Neural Network (NN) that predicts whether a configuration will crash, its expected performance, and the performance uncertainty of these predictions.
It consists of two branches: the \textit{prediction branch}, which is a conventional NN that predicts permutation crash probability and performance, and the \textit{uncertainty branch}, which is a Radial Basis Function NN that estimates the uncertainty of the predicted performance.

In the following, we note vectors in bold (\textbf{x}), matrices in capitals (\textit{X}), and scalars in lower-case (\textit{$\alpha$}).
We can define the DTM as a function $F(\textbf{x}) \rightarrow (\hat{k}, \hat{y}, \hat{\sigma})$ that maps a configuration permutation $\textbf{x}$ into the crashing probability $\hat{k}$, the expected performance $\hat{y}$, and the predicted uncertainty $\hat{\sigma}$.
Each configuration can be divided as $\textbf{x} = (\textbf{x}^k, \textbf{x}^n)$, with $\textbf{x}^k \in \mathbb{N}^k$ the subset of discrete/categorical parameters (e.g., Nginx's \texttt{DEFAULT\_QDISC} parameter can be a string "pfifo", "bfifo", etc.) and $\textbf{x}^n \in \mathbb{R}^n$, for the continuous/integer parameters.
$F(\textbf{x})$ can be divided into its uncertainty branch $F^u(\textbf{x}) \rightarrow (\hat{k}, \hat{y})$ and prediction branch $F^p (\textbf{x}) \rightarrow (\hat{\sigma})$.

$F^p$ is a conventional feedforward deep NN~\cite{JMLR:v15:srivastava14a, pmlr-v202-mao23b, KendallG17}.
It consists of a sequence of dense layers with \emph{Rectified Linear Unit} (\textit{ReLU}) activation functions and dropout~\cite{JMLR:v15:srivastava14a} layers.
Its last layer outputs performance and crash predictions.

$F^u$ (pink area in \cref{fig:deeptune_model}) is specifically designed to estimate uncertainty.
It consists of a stack of Gaussian Radial Basis Function (RBF) layers~\cite{Lee1999}, each consisting of neurons $\phi$ that contain centroids $\textbf{c} = (c_1, c_2, c_{\dots}$).
These centroids, learned throughout the training process, can be interpreted as learned features (or \emph{prototypes}) from the dataset.
Each RBF layer is parallel to a layer of the prediction branch.
It takes as input $\textbf{z}$, the concatenation of the features/latents output by the previous layer. The activation value $\phi$ is computed as:
\begin{equation}
\phi(\textbf{z}) = \exp\left(-\frac{\|\textbf{z} - \textbf{c}\|^2}{2\gamma^2}\right)
\end{equation}

$\gamma$ is a smoothing parameter that controls how flat the activation curve of each neuron is.
This parameter should be empirically fit: we find that a $\gamma$ value of 0.1 is appropriate if input features are z-score normalized.
The motivation behind this design is to be robust to outliers or totally new samples: the response of each neuron depends on the distance to the learned centroid or data prototypes (represented by the norm $\|\dots\|$), hence when an outlier appears, the output is low.

\noindent \textbf{Training the DTM.} We train the DTM end-to-end by minimizing the loss function $\mathcal{L} = L_{\text{CCE}} + L_{\text{Reg}} + L_{\text{Cham}}$.
We import these three components from prior works:

% Note from Hugo: I removed the equations as they will not be understood by the systems folks, and had typos anyways.
\begin{itemize}
\item The categorical cross-entropy loss $L_{\text{CCE}}$~\cite{pmlr-v202-mao23b} enables the DTM to identify which permutation might crash through learning historical data. 
%\begin{equation}
%L_{\text{CCE}} = -\frac{1}{N} \sum_{i=1}^{N} \textbf{c}_{i} \log(\hat{\textbf{c}}_{i})
%\end{equation}
%With $N$ is the number of samples, and $\textbf{c}_i$ and $\hat{\textbf{c}}_i$ are the one-hot encoding of the ground truth and predicted value of the $i\text{th}$ sample.

\item The regression loss with uncertainty $L_{\text{Reg}}$~\cite{KendallG17} addresses the challenge of performance prediction while quantifying uncertainty.
Initially contributed for computer vision tasks~\cite{KendallG17}, we can use it to allow the DTM to both estimate the expected performance profile of a given permutation and provide an estimation of the expected error of the prediction. This enables better choices in the next permutation sampling policy.
%\begin{equation}
%L_{\text{Reg}} = \frac{1}{2N} \sum_{i=1}^{N}  \frac{1}{2\hat{\sigma_i}^2} \left \| y_i - \hat{y}_i \right \|^2 + \frac{1}{2} \text{log} (\hat{\sigma}_i^2)
%\end{equation}

\item The regularization loss $L_{\text{Cham}}$~\cite{fangsg17} for RBF layers enables the DTM to learn centroids $\textbf{c}$ that fit the training data.
%\begin{equation}
%L_{\text{Cham}} = \frac{1}{N} \sum_{i=1}^{N}  \min_{\mathbf{x} \in X} \left \| \boldsymbol{x}_i - \mathbf{z}_i \right \|_2^2 + \min_{\mathbf{z} \in Z} \left \| \boldsymbol{x}_i - \mathbf{z}_i \right \|_2^2
%\end{equation}
Intuitively, $L_{\text{Cham}}$ is the \emph{Chamfer}~\cite{fangsg17} distance, initially contributed in computer vision to compute the distance between two point clouds~\cite{Barrow1977}.
Here, it computes the distance between the centroids and the input data; minimizing it effectively distributes centroids such that they fit the training data distribution.
\end{itemize}

While $L_{\text{CCE}}$, $L_{\text{Reg}}$, and $L_{\text{Cham}}$ exist from prior works in other fields~\cite{pmlr-v202-mao23b, KendallG17, fangsg17, Barrow1977}, we contribute a new way to use them in $\mathcal{L}$ to solve our problem.

\noindent \textbf{The scoring function.} We must strike a balance between exploration and exploitation when selecting the next configuration to try.
We thus contribute a scoring function $\text{sf}(\textbf{x}, X)$ that leverages both the dissimilarity of a sample \textbf{x} compared to known sample points $X$, and their estimated error or uncertainty $\hat{\sigma}$ to prioritize exploration in under-explored regions, while exploiting areas with greater expected improvements.

Specifically, the scoring function starts by calculating the dissimilarity $\text{ds}(\textbf{x}, X)$ between the candidate point $\textbf{x}$ in the search space and known sample points $X$.
This dissimilarity metric accounts for the diversity of the samples and indicates regions that have not yet been thoroughly explored:
\begin{equation}
\label{eq:ds}
    \text{ds}(\textbf{x}, X) = 1 - \frac{1}{\left (1+\left || \textbf{x} - X \right||_2^2 \right )}
\end{equation}

% TODO Daniel: we need to give a formula for s(), just like we do for ds(). Then we need to explain why we need this new formula compared to previous approaches (why it's necessary). It's also not clear what "standard scores" means.

% We can just remove this paragraph. It doesn't have sense, and it doesn't add useful information :)
% We then incorporate the standard scores of the existing samples, denoted as $s(\textbf{x}, X)$, reflecting the performance and uncertainty associated with each point.
% Multiple approaches for $s(\textbf{x}, X)$ exist~\cite{SnoekLA12, MockusPRSZ17, JonesSW98, SMAC3, Hernandez-LobatoHG14}; we settle for\dots % TODO Which approach do we take?

%Multiple approaches for $s(\textbf{x}, X)$ exist, including the Expected Improvement \cite{SnoekLA12, MockusPRSZ17, JonesSW98}, the Probability of Improvement \cite{SMAC3}, or more recently the Predictive Entropy \cite{Hernandez-LobatoHG14}.

We calculate the final score by combining the sample dissimilarity and the predicted sample uncertainty, represented by $F^u(\textbf{x})$ (which maps $\textbf{x}$ to $\hat{\sigma}$, see Figure~\ref{fig:deeptune_model}), using a predefined weight $\alpha$:

\begin{equation}
\label{eq:scoring_function}
    \text{sf}(\textbf{x}, X) =  \alpha \text{ds}(\textbf{x}, X) + (1-\alpha) F^u(\textbf{x}).
\end{equation}

Our experiments indicate that setting \(\alpha\) to 0.5 strikes an effective balance between exploitation and exploration.

The justification behind our choice of loss and scoring functions is as follows: the loss function trains the system to simultaneously identify invalid (crashing) configurations and predict performance, while the scoring function uses those predictions to strategically choose new configurations that balance exploration vs. exploitation. The RBF branch adds a mechanism that allows \algo to estimate its confidence/uncertainty.

Our implementation of \algo optimizes for a single metric at a time.
However, it can be extended to handle multiple metrics by adding additional output layers to $F^p$ and $F^u$. This modification allows the DTM to make predictions for multiple targets simultaneously. During the scoring phase, we apply equation \ref{eq:scoring_function} to each target metric to obtain individual scores. Then, we calculate a representative score for each permutation sample by taking a weighted average, or using another aggregation method, of these individual scores.

\subsection{Transfer Learning}

\begin{figure}
    \centering
    \includegraphics[width=.9\linewidth]{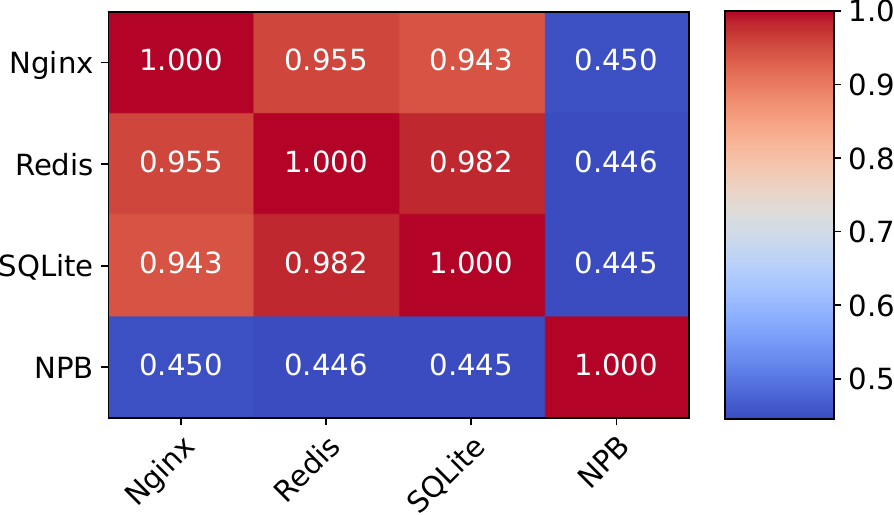}
    \caption{Cross-similarity matrix: a value close to 1 at the intersection of a column and a row means that the performance of the applications is impacted by similar parameters.}
    \label{fig:tl-matrix}
\end{figure}

By default, each process of optimizing for one application and metric starts with a blank model, and parameters impacting performances/crashing must be re-learned from scratch.
Transfer learning~\cite{hosna2022transfer} consists of pre-training a model on one application, and re-using it to speed up the optimization process for another related application.
The intuition behind transfer learning lies in the similarity of features between tasks: when applications share characteristics and the metrics to optimize towards are similar, it is probable that a model pre-trained on one application will be useful for the other.
In other words, the performance of two applications may be sensitive to the variation of the same configuration parameters.
For example, both Redis and Nginx are network-intensive: when an instance of DeepTune is trained on one, the subset of parameters and values that matter for the other (e.g., network stack parameters) has already partially been identified, speeding up the search.
Conversely, that particular instance will be ineffective with NPB which is CPU-/memory-intensive.

To confirm that hypothesis, we build a cross-similarity matrix (\Cref{fig:tl-matrix}) to assess the similarity/differences between the most performance-impactful kernel configuration options for the four applications considered in the evaluation (see \S\ref{sec:evaluation} for details).
To create that matrix, we first collect 2,000 random Linux configurations for each application.
Then, we use a feature importance algorithm~\cite{breiman2001random} to determine the importance of each configuration option in predicting performance.
Finally, we treat the importance scores as vectors and compute the Euclidean-norm distance between them.
The parameters with the highest performance impact are similar for Nginx, Redis, and SQLite, which are all system-intensive.
Redis is closer to SQLite than it is to Nginx, which is unsurprising as Redis and SQLite are both databases.
NPB, on the other hand, exhibits sensitivity to different parameters due to its compute and memory-intensive nature.
This suggests that transfer learning can be effective for certain classes of applications in \sys.

\subsection{Defining the Exploration Space}
\label{sec:defining-the-space}

\sys requires a description of the configuration space to start the exploration process (the \emph{job file}, as mentioned in \S\ref{subsec:overview-bench-platform}).
That description includes the list of configuration parameters, their types, and the possible values they can take.
Modern systems, such as Linux, have complex configurations, so completely describing the configuration space is challenging.
Although some information can be statically obtained for compile- and runtime parameters (e.g. by analyzing \texttt{Kconfig} files and kernel command line parameter descriptions~\cite{KERNEL_CMDLINE}), it is not the case for runtime parameters: these can be poorly or not documented, and for many parameters the range of valid values cannot be statically determined, because it depends on runtime aspects (e.g., the amount of available RAM).

We developed the following heuristic approach to determine the Linux configuration space.
Linux offers multiple runtime configuration options through virtual file systems, e.g. \texttt{/proc/sys}, \texttt{/sys}.
We first determine all configuration options by booting a VM with the version of Linux under investigation, and listing writable files in these paths.
For each writable file, we read it and assume the value returned corresponds to the default value for the corresponding configuration options.
Next we determine the type of the option by checking the type of the default value.
If it is a number and equals 0 or 1, we assume the option is boolean.
If it is neither 0 nor 1, we treat it as an arbitrary integer.

Finally, we estimate the range of possible values for the option by scaling up and down the default value several times by a high factor (10) and attempting to set the option to these new values by writing to the corresponding pseudo file.
If the write operation succeeds and the VM does not crash, we consider the new value to be in the valid range for that option.
The exploration of configuration values is left intentionally coarse, as it will be the task of \sys to find performance improvements.
Note that this technique excludes configuration parameters that are not numbers (e.g., strings), for which valid values would be very hard to determine automatically.
There are very few non-numeric runtime parameters (see Table \ref{tab:linux-parameter-types}), and for these, we call back to manual exploration when necessary.
\sys will explore categorical parameters one by one, with no assumption of relationship (e.g., linear) between the values they can take.
String parameters will not be explored beyond the values that can be automatically extracted.
Beyond Linux, this approach should generalize easily to other OSes.
We argue that more formal specifications of kernel parameters would be desirable in the future, taking inspiration from the system call interface definitions contributed by the fuzzing community~\cite{SYZLANG}.

\subsection{Integration in Deployment Workflows}

\paragraph{Practicality: Integrating \sys in Common Deployment Workflows.}

We envision \sys to be used in the testing/evaluation phases of an application's development.
Equipped with a workload and a machine representative of deployment conditions, the system optimizes the kernel's configuration for the application running in these conditions.
At that stage, if a configuration identified by the platform is to be deployed to production, an engineer should review the deployment to ensure that it meets production requirements.
While \sys can check that the system is functional as part of its benchmarking step, like any AI tool it cannot exonerate a site reliability engineer (and, typically, a CI pipeline) from checking the configurations it pushes to production.
If a configuration does not meet production requirements, users can further constrain the exploration to weed out that part of the space:

\begin{compactitem}
  \item \emph{Constraining the parameters covered by the search.}
Users can optionally specify a set of parameters which should take fixed values and will not be varied by \sys's search process.
This is for example useful to ensure that relevant security options (ASLR, etc.) are not disabled by \sys.
\sys can also be instructed to favor varying certain parameter types (compile-time/boot-time/runtime), which is useful, e.g., when the kernel to optimize cannot be rebooted.

  \item \emph{More comprehensive benchmarks.}
Users can also extend the benchmarking tool used by \sys (purple in Figure~\ref{fig:deeptune_model}) to check additional functionalities of the deployment (e.g., run a test suite).
If the check fails, \sys will learn the kernel configurations that cause the misbehavior.
\end{compactitem}

Overall, our design builds on a simple assumption: while users need not have expert knowledge on OS performance optimization, we assume that they can check if a deployment meets their production requirements.

\paragraph{Security Considerations}

In security-sensitive scenarios, relying without expert knowledge on a fully automated framework raises legitimate safety concerns.
The threat model here is as follows: without awareness of security-relevant options, \sys may overlook, disable, and optimize out such options, leading to kernel configurations which may be exploited in the field if deployed as is.

As mentioned previously, we can enable a security-aware search mode by fixing security-critical parameters to safe values.
In a context where we envision no particular expertise from the user, this means that we assume the capacity to identify such important security-related options.
We believe that this assumption is justified: commonly-used security parameters can be identified by \sys's developers, requiring no effort/expertise from the user.
It is also safe to assume that the user is aware of any non-standard security parameter that should be set to a particular value, otherwise they would not be able to run their application securely in the first place.
Finally, as mentioned above, like with any other automated tool, \sys's output should be checked by a reliability engineer before being pushed to production.

\paragraph{Sensitivity to Workload and Hardware}
\sys specializes a kernel configuration for a particular application running on a particular platform and processing a particular workload.
Similarly to most performance evaluation works, a change in workload or hardware requires rerunning the evaluation to obtain accurate results for the software and hardware considered.
\sys could be extended to predict performance for hardware/workloads that are different from those evaluated, using cross-workload~\cite{WORKLOAD_PREDICTION} and cross-platform~\cite{CROSS_PLATFORM} performance estimation methods e.g., from the heterogeneous computing literature~\cite{AMP, HEXO}.
We scope out that objective as future work.

%% file: 04-evaluation.tex
\section{Evaluation}
\label{sec:evaluation}

This evaluation answers the following questions:
\begin{itemize}
    \item Given the vast configuration space offered by OSes, how quickly can \sys converge on a specialized configuration for a given application?
    For a given iterations/time budget, what is the best configuration that can be found by \sys?
    How do these results compare to baselines and competitors? (\S\ref{sec:eval-perf})

    \item When \sys pre-trains a model on an application, what are the benefits of applying transfer learning and using that model to speed up the search of a specialized configuration for other applications? (\S\ref{sec:eval-tl})

    \item How good is \sys at predicting crashes? (\S\ref{sec:eval-crashes})

    \item Beyond Linux and performance, can \sys be applied to different OSes and other metrics? (\S\ref{sec:eval-misc})

\end{itemize}

% Pierre: need to put this pic here so that it is close to the text referencing it
\begin{figure*}
    \centering
    \begin{subfigure}{0.49\textwidth}
        \centering
        \includegraphics[width=\textwidth]{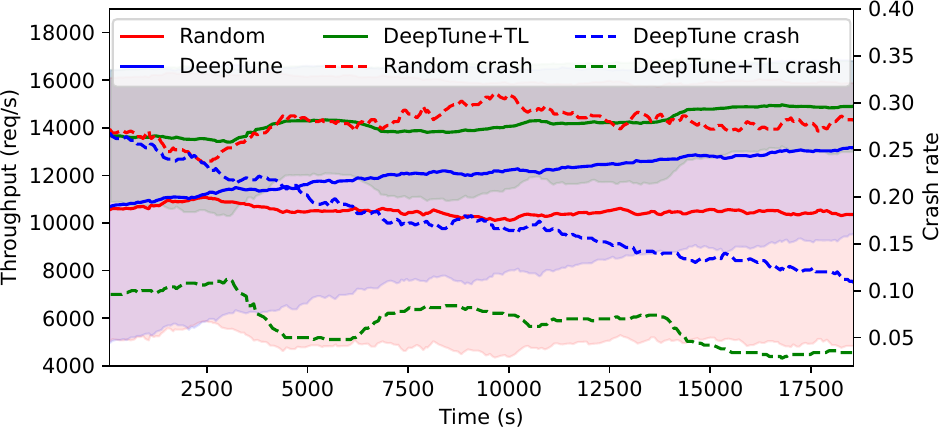}
        \caption{Nginx (performance: throughput, \textbf{higher is better}).}
        \label{fig:mean and std of net14}
    \end{subfigure}
    \hfill
    \begin{subfigure}{0.49\textwidth}
        \centering
        \includegraphics[width=\textwidth]{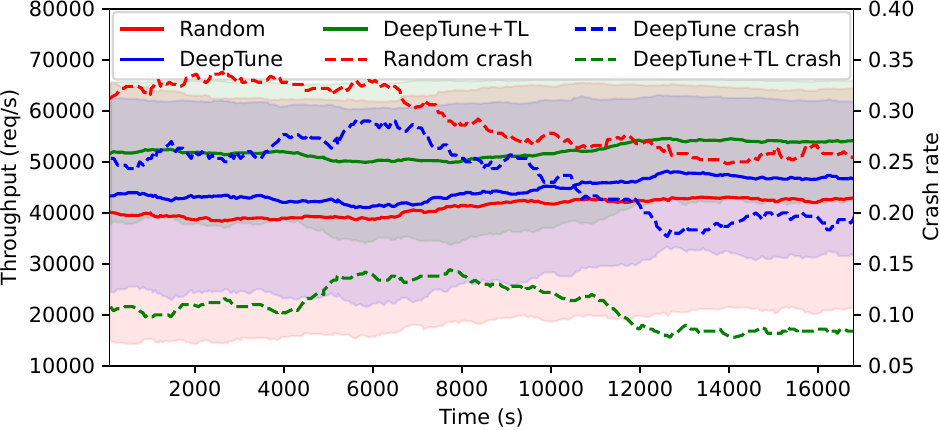}
        \caption{Redis (performance: throughput, \textbf{higher is better}).}
        \label{fig:mean and std of net24}
    \end{subfigure}
    \vskip\baselineskip
    \begin{subfigure}{0.49\textwidth}
        \centering
        \includegraphics[width=\textwidth]{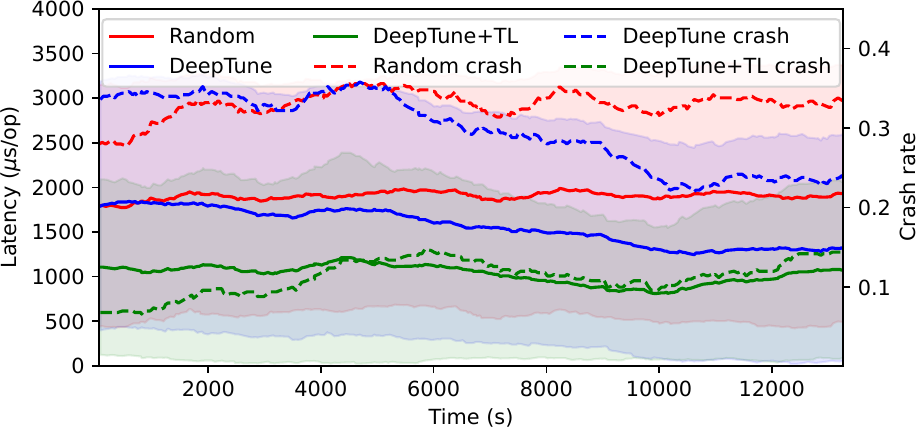}
        \caption{SQLite (performance: operation latency, \textbf{lower is better}).}
        \label{fig:mean and std of net34}
    \end{subfigure}
    \hfill
    \begin{subfigure}{0.49\textwidth}
        \centering
        \includegraphics[width=\textwidth]{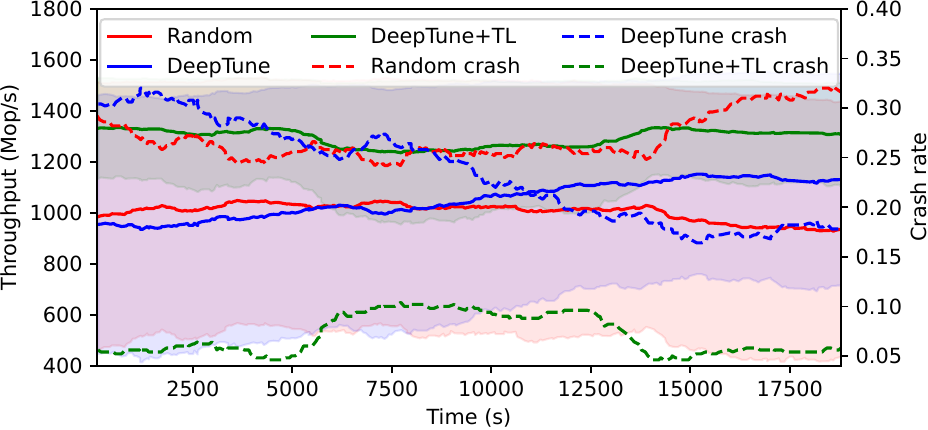}
        \caption{NPB (performance: Mop/s, \textbf{higher is better}).}
        \label{fig:mean and std of net44}
    \end{subfigure}
    \caption{Evolution of performance of configurations (solid lines, results of 5 runs smoothed for readability, with the area behind each curve representing the spread) found by a search session of 250 iterations for Nginx, Redis, SQLite, and NPB with \sys, without and with transfer learning (TL, trained with Redis), and random search. Dashed lines show the crash rate.} 
    \label{fig:perf-general}
\end{figure*}

We select a set of popular applications: the Nginx~\cite{NGINX} web server and the Redis~\cite{REDIS} key-value store (both network-intensive), the SQLite~\cite{SQLITE} database management system (storage-intensive), and the NAS Parallel Benchmarks~\cite{NPB} (NPB, CPU- and memory-intensive).
Nginx is benchmarked with \texttt{wrk}~\cite{WRK}, Redis with \texttt{redis-benchmark}~\cite{REDIS_BENCHMARK}, and in both cases we optimize to maximize throughput.
SQLite is benchmarked with LevelDB's SQLite3 benchmark~\cite{LEVELDB_SQLITE3}, in which a high number of SQL \texttt{INSERT} operations are issued, and we aim to minimize the average execution time of each operation.
We use the OpenMP version of NPB and select a mix of CPU- and memory-intensive programs: FT, MG, CG, IS (running the entire suite is too long) with size classes S, W, A, and B.
In one run we execute all programs for each size class and aggregate the number of operations per second reported by each benchmark, which is the metric we aim to maximize.
We run all experiments on a dual-socket server with 2 Intel Xeon E5-2697 v2 (2x24 cores at 2.70 GHz, 128 GB of RAM), running Debian 10, and configured for high and stable performance: isolcpus on core 0-1, hyperthreading and ASLR disabled, performance CPU governor.
Redis and SQLite run on 1 core because of their single-threaded nature, while Nginx and NPB run on 16 cores.
The server used exposes 2 NUMA nodes, but we restrict the experiments to a single one to avoid NUMA skewing performance measurements.
To avoid any disturbance due to experiment co-location, all test configurations are benchmarked one after the other and there is no experiment co-location.

Unless otherwise stated (\S \ref{sec:eval-tl}), with \sys transfer learning is disabled and each search process starts from scratch with a blank model.
As a baseline we select random search~\cite{RANDOM_SEARCH}, an approach well-known to give good results on large exploration spaces, such as the ones we target.
This method explores the design space by continuously generating unique configurations with random values for each parameter.
We omit comparing with grid search, which is well-known to be inferior to random search on the large configuration spaces we target.
We also show that \sys performs better than Bayesian optimization~\cite{BAYESIAN_OPT}, a common technique used to optimize long-running black-box functions, which uses a probabilistic model to balance trying out previously unexplored parameters and focusing on parameters known to impact performance.
We also demonstrate \sys's superiority to Unicorn~\cite{UNICORN}, a closely related work focusing on the optimization of OS and application configurations using causal inference.
As we demonstrate, Bayesian optimization and Unicorn do not scale to the large configuration spaces targeted by \sys (e.g. Linux configuration), hence we compare to them on smaller exploration spaces.
We also present how \sys synergizes with compile-time, dynamic analysis based optimizers, like Cozart~\cite{COZART}, a related work that leverages dynamic analysis to significantly reduce the number of Linux kernel configuration options, resulting in a much smaller footprint and exploration space.
We show that configurations generated by Cozart are veritable baselines for \sys to optimize upon through run-time options. 
Finally, for all experiments the initial OS configuration used to kickstart each search process is defined randomly.

\subsection{Performance of the Configuration-Space Search}
\label{sec:eval-perf}

\paragraph{OS Configuration Specialization.}

We run \sys to specialize the Linux kernel towards performance, for each target application.
We use Debian 10's kernel v4.19 (a long-term support version).
For these performance-based experiments, we configure \sys to favor exploration of runtime parameters.
We run both \sys and random search for 250 iterations (250 parameters explored, which takes 3.5 to 5.2 hours depending on the application), and compare the performance of the configurations found.

\Cref{tab:default-conf} shows the performance of the best configurations found by \sys, along with the average time taken to find them and the performance of the default Lupine Linux~\cite{LUPINE} configuration.
Lupine is a Linux kernel specialized for general purpose performance, i.e., not for any particular application.
The configuration found for Nginx is 24\% faster, showing that \sys can successfully find configurations that perform better than default.
For NPB the improvement is only of 2\%.
Since the benchmark is mostly CPU- and memory-intensive and does not request any system functionality, the OS configuration has close to no impact on its performance.
The best configuration found for SQLite also does not improve performance, which seems to indicate that the default configuration is already highly efficient for this scenario.

Figure~\ref{fig:perf-general} presents the performance of configurations found by each approach.
Dashed lines represent the crash rate, with 1 corresponding to a crash every iteration.
The performance of the configurations found by \sys is at the beginning of the search process similar to that of random search.
After a certain number of iterations, \algo's neural networks learn important parameters and how to use them efficiently, causing the configurations' performance to outperform that of random.
For example, for Nginx, after 250 iterations the smoothed throughput is more than 20\% higher than that of random search.
Unlike random's relatively consistent crash behavior, Trailblazer's crash rate decreases over time, as it learns to avoid configurations likely to crash: e.g., with Nginx the crash rate goes from 0.3 to 0.1 after 250 iterations.

\begin{table}
    \center
    \footnotesize
    \caption{Best-performing configurations found by \sys applied to Linux v4.19 after 250 iterations.}

\begin{tabular}{|c|p{3.2em}|p{2.5em}|p{2em}| c| c | c|}
    \hline
    \multirow{2}{*}{{\bf App.}} & \multirow{2}{*}{\shortstack[c]{{\bf Lupine}\\{\bf Linux}}} &\multirow{2}{*}{\shortstack[c]{\bf Way-\\\textbf{finder}}} & \multirow{2}{*}{\shortstack[c]{{\bf Perf.}\\{\bf unit}}} & \multirow{2}{*}{{\shortstack[c]{\bf Relative\\\bf Perf.}}} & \multicolumn{2}{c|}{\shortstack[c]{{\bf Avg. time}\\{\bf to find}}}   \\
     \cline{6-7}
    & & & & &  {\bf No TL} & {\bf TL} \\
    \hline
    Nginx  & \raggedleft 15731 & \raggedleft 19593 & req/s     & \raggedleft 1.24x & \raggedleft 415s & 92s \\
    \hline
    Redis  & \raggedleft 58000 & \raggedleft 66118 & req/s     & \raggedleft 1.14x & \raggedleft 312s & 69s \\
    \hline
    SQLite & \raggedleft 284   & \raggedleft 284   & $\mu$s/op & \raggedleft 1x & \raggedleft 248s & 76s \\
    \hline
    NPB    & \raggedleft 1497  & \raggedleft 1522  & Mop/s     & \raggedleft 1.02x & \raggedleft 243s & 76s \\
    \hline
\end{tabular}
    \label{tab:default-conf}
\end{table}

\paragraph{Scaling to Large Configuration Spaces and Long Search Processes.}

\begin{figure}
    \centering
    \includegraphics[width=0.40\textwidth]{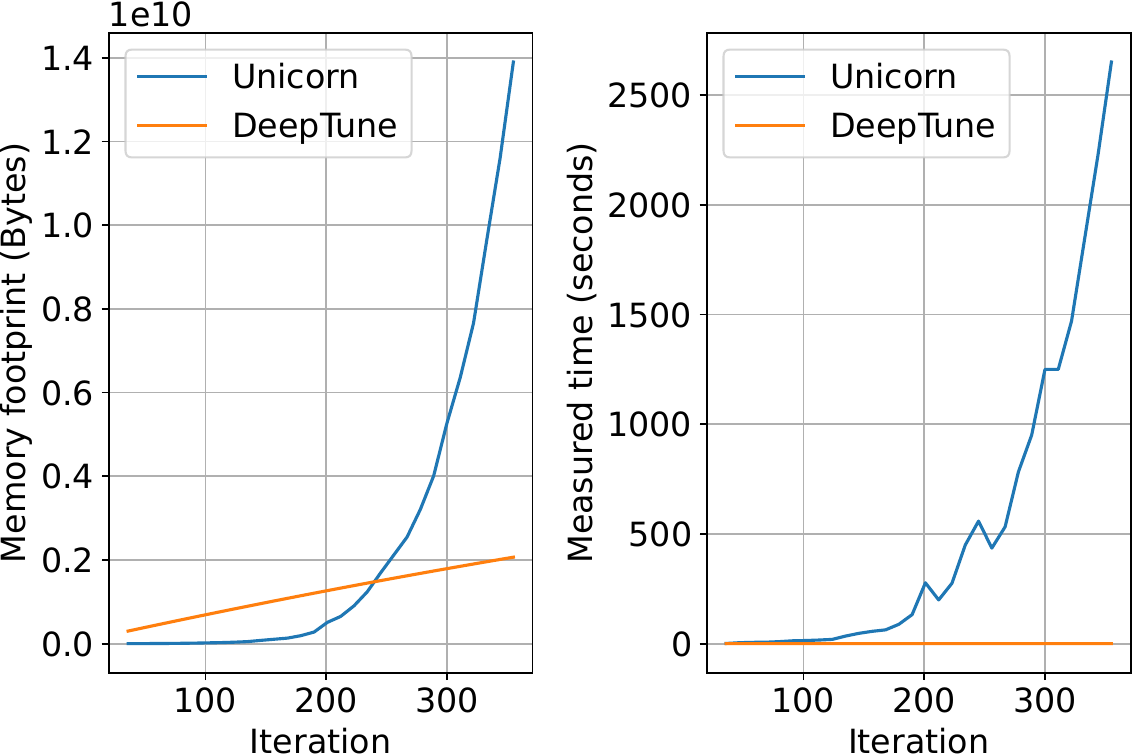}
    \caption{Evolution of the memory consumption (left) and algorithm execution time (right) of \algo vs. Unicorn over a run of the search process.}
\label{fig:Scalability}
\end{figure}

Here we compare the scalability of \sys with a competitor, Unicorn~\cite{UNICORN}.
As Unicorn cannot scale to the size of Linux's configuration, we create a synthetic dataset with known local and global maximas to increase the likelihood of convergence.
The dataset has a total number of parameters that match those used in the original Unicorn paper~\cite{UNICORN}.
We measure the evolution of the execution time and peak memory consumption (using Python's tracemalloc~\cite{TRACEMALLOC}) of each iteration for both algorithms.
The results are presented on \Cref{fig:Scalability}.
As one can observe, both the execution time and memory consumption of Unicorn increase exponentially while the algorithm iterate.
These limitations come from Unicorn's design choice of the causal analysis algorithm, which typically has $O(n^4)$ complexity, making it a poor fit for the large search spaces of modern OS configurations.
In contrast, the complexity of \algo grows linearly $O(n)$ in both time and memory consumption.

Due to these scalability issues, we are unable to compare to Unicorn in the other experiments presented in this evaluation.
Indeed, these target design spaces which are several orders of magnitude larger than what Unicorn can handle.

\paragraph{Search Loop Execution Time.}

\begin{figure}[t]
    \centering
    \includegraphics[width=0.40\textwidth]{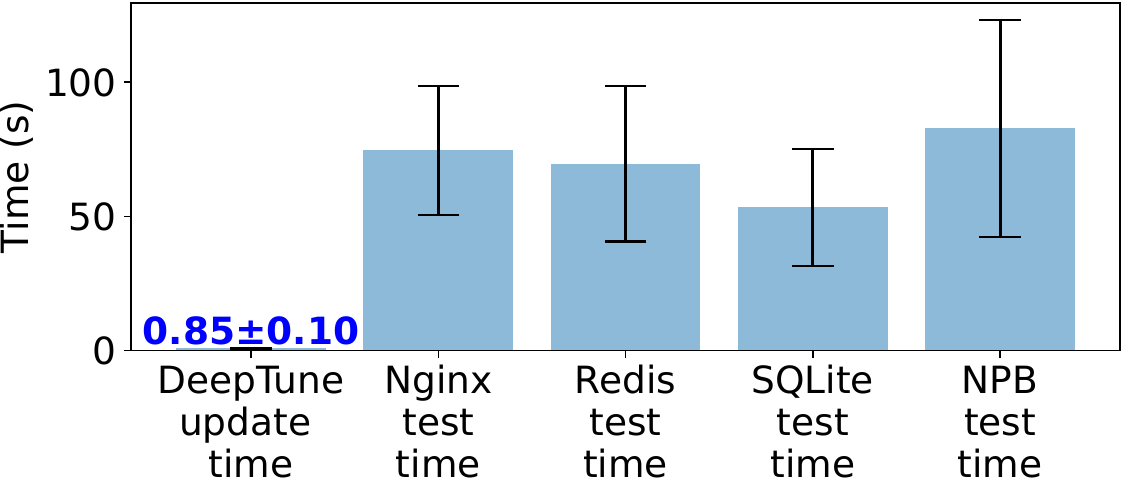}
    \caption{Average update time of \algo vs. the average test time for different applications.}
    \label{fig:exploration-loop}
\end{figure}

We measure the average execution time for evaluating a single configuration to understand \sys's performance.
\Cref{fig:exploration-loop} presents the execution time, broken down into the time spent running \algo to decide what configuration to evaluate, and the time spent evaluating the configuration, i.e. booting the kernel, launching the target application, running the experiment, etc.
Evaluating a configuration dominates the search process: it takes on average 60-80~s depending on the application, and varies widely due to the performance impact of the configurations evaluated.
Conversely, the execution time of an iteration of \algo takes less than a second, showing that the bottleneck is in the evaluation of configurations and not in the search algorithm.

\paragraph{High-Impact Configuration Parameters}
\label{sec:impactful-parameters}
We queried the models learned by \algo to assess \sys's ability to identify parameters with the high impact on performance, including some that have been previously identified by experts.
Here we focus on Nginx for space reasons, and because it is the target of many performance tuning guides.

Regarding top parameters impacting performance \emph{positively}, while Trailblazer identifies parameters that are well-documented in tuning guides, it also uncovers other important configuration options that are not mentioned.
For example, it identifies options such as the maximum number of connections that can be queued in the TCP/IP stack backlog per socket (\texttt{net.core.somaxconn}), the default size of sockets receive buffer (\texttt{net.core.rmem\allowbreak\_default}), and the TCP keepalive time (\texttt{net.ipv4.tcp\_keep\allowbreak alive\_time}).
These have been documented in Nginx/network performance tuning guides as high-impact parameters~\cite{GUIDE_SOMAXCONN, GUIDE_SOMAXCONN2, GUIDE_RMEM, GUIDE_KEEPALIVE}.
However, \sys also identifies parameters impacting performance in a less intuitive way, e.g. the frequency at which memory statistics such as those reported by vmstat are computed (\texttt{vm.stat\_interval}).
As mentioned in tuning guides~\cite{GUIDE_STAT_INTERVAL}, reducing it helps in latency sensitive scenarios.
Hence, \sys can automatically pinpoint and optimize for non-obvious parameters that have been identified by experts as impacting performance.
Regarding top parameters \emph{negatively} impacting performance, \sys identifies several options leading to significant performance degradation, for example setting high levels of kernel verbosity (\texttt{printk}), delaying kernel logs (\texttt{printk\_delay}) and enabling block I/O debugging (\texttt{vm.block\_dump}).
Logging and debugging are well-known to impact Nginx's performance~\cite{WAYFINDER}, and once again this demonstrates \sys's ability to identify documented bottlenecks.

\subsection{Transfer Learning Efficiency}
\label{sec:eval-tl}

To assess the efficiency of transfer learning, we trained a model with \algo on Redis for 250 iterations (taking 4.6 hours), and evaluated that model's efficacy at finding specialized configurations for the other 3 applications considered in this evaluation.

The results are presented on \Cref{fig:perf-general}, with the curves labeled ``TL'' in the legend showing the performance of the configuration found and the crash rate of the transferred model.
Using transfer learning, the configuration performance is consistently higher compared to starting the exploration process with an untrained model: for example, the first configuration found at the beginning of the process for Nginx has 1.33$\times$ higher performance with transfer learning than without transfer learning or with random search.
The crash rate is also generally much lower with transfer learning, being below 10\% in most cases.
\Cref{tab:default-conf} takes another glance at the efficiency of transfer learning.
It presents, in its last two columns, the time taken to reach a specialized configuration with and without transfer learning.
The time savings brought by the technique are important: the search is sped up by a factor between 4.5$\times$ (Nginx) and 3.2$\times$ (NPB).
Please note that, as the curves presented on \Cref{fig:perf-general} are averages of multiple runs, it is unsurprising that the curve for \algo + TL does not start exactly where the curve for \algo without transfer learning ended in \Cref{fig:mean and std of net24}.

\subsection{Crash Prediction}
\label{sec:eval-crashes}

Using random exploration, a high number of configurations (about one third) fail at runtime, which is a significant waste of time.
\sys also faces failures, however its ability to learn the configuration parameters that are likely to trigger crashes means that it can avoid many of them.

\begin{table}
    \center

    \caption{Base prediction accuracy (1 being 100\% accuracy) of \algo. \textit{Failure acc.} and \textit{Run acc.} shows the prediction accuracy for a failure and an application running (non-failure) events. MAE is the normalized mean absolute error.}

    \footnotesize
    \begin{tabular}{|c | c| c | c|}
    \hline
     {\bf Application} & {\shortstack[c]{{\bf Failure}\\{\bf accuracy}}} &  {\shortstack[c]{{\bf Run}\\{\bf accuracy}}} & {\shortstack[c]{{\bf Performance prediction}\\{\bf normalized MAE}}}   \\
      \hline
      Nginx  & 0.796 & 0.397  & 0.273 \\
      \hline
      Redis  & 0.789 & 0.310  & 0.361 \\
      \hline
      SQLite & 0.742   & 0.456  & 0.112 \\
      \hline
      NPB    & 0.755  & 0.455   & 0.359 \\
      \hline
    \end{tabular}

    \label{tab:crash-prediction}
\end{table}

\Cref{tab:crash-prediction} presents the accuracy of \algo when predicting that a given configuration will fail (failure accuracy) and when predicting that a given configuration will execute successfully (run accuracy).
It also presents the normalized mean absolute error when predicting the performance of a configuration.
As one can observe, the failure accuracy is high, being between 75\% and 80\%, allowing \sys to avoid a large number of crashes.
The run accuracy is low and ranges between 0\% to 36\%, hence we rely on failure accuracy to estimate the probability of a configuration failing, to determine if it is worth or not to evaluate that configuration.

% TODO: It would be really nice if I could get a list of parameters that lead to runtime crashes. We can have a quick look and mention them in Section 4.3. It would further help eliminating the feeling of "this thing is magic".

\subsection{Varying OSes and Target Metrics}
\label{sec:eval-misc}

\paragraph{Application to Other OSes: Unikraft.}

To demonstrate \sys's application to OSes other than Linux, we now apply \sys to the Unikraft~\cite{UNIKRAFT} library OS.
We compile an Nginx Unikraft image and optimize its configuration to maximize request throughput using \sys.
Unikraft exposes a configuration space that is much smaller than that of Linux, making it is possible to compare with Bayesian optimization: we explore 33 configuration parameters (10 Nginx application-level parameters, and 23 Unikraft OS parameters), yielding a search space of $3.7 \times 10^{13}$ permutations.

\begin{figure}
    \includegraphics[width=0.45\textwidth]{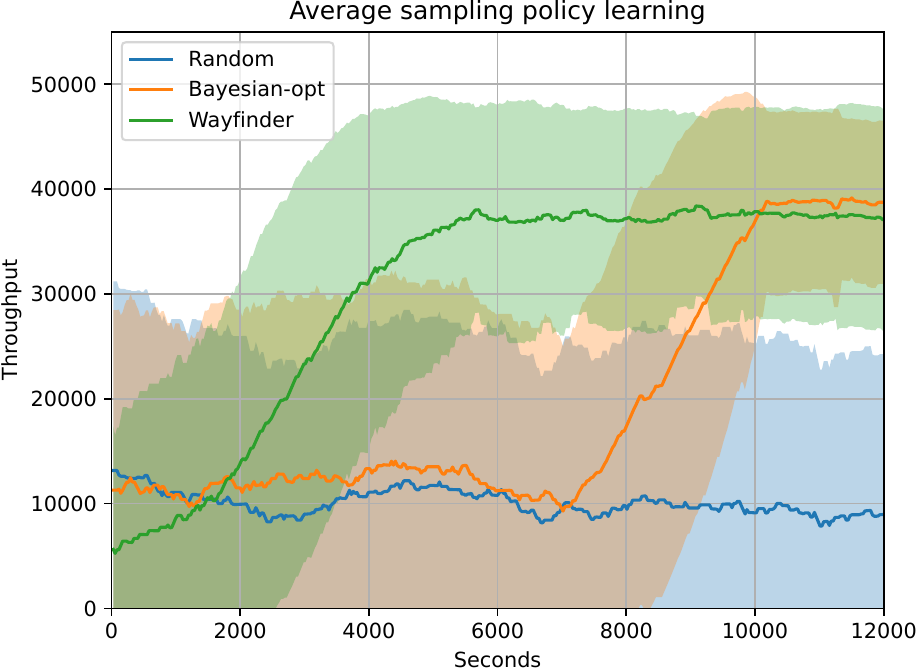}
    \caption{Evolution of the performance of configurations found by \sys, random search, and Bayesian optimization for Nginx running on the Unikraft unikernel. Results of 5 runs, smoothed for readability.}
    \label{fig:unikraft-nginx}
\end{figure}

We launch the exploration process with a time budget of 3 hours.
The performance of the configurations found is presented in \Cref{fig:unikraft-nginx}.
As one can observe, \sys quickly converges on a specialized configuration, reached after 100 minutes.
Bayesian optimization takes more than 160 minutes to reach configurations that perform similarly.
With that time budget, we also observe that random search is not able to find high-performance configurations.
\Cref{fig:unikraft-nginx} clearly displays three phases in the behavior of \algo.
\algo first slowly uncovers configurations with better performance.
After about 25 minutes it picks up an impactful set of parameters and enters an exploitation phase where it focuses on the parameters to rapidly increase performance.
This phase ends after about 100 minutes, at which points \algo goes back to exploration.

% TODO Here if only we could say what are these parameters... It would REALLY strengthen this part.

Note that the performance improvement brought by top configurations is significantly higher than that observed for Linux (see \Cref{fig:perf-general}).
This can be explained by the fact that Unikraft is a unikernel~\cite{MIRAGE_ASPLOS} offering low-latency user/kernel transitions which, under the right configuration, speed up latency-sensitive system intensive workloads significantly~\cite{UNIKRAFT}.

\paragraph{Application to Other Metrics: Memory Footprint.}

To assess \sys's suitability to optimize beyond performance, we measure its efficiency to find Linux kernel configurations with small memory footprint.
This metric has been the target of several past works, since it is crucial in certain domains, such as lightweight virtualization~\cite{XCONTAINER, LUPINE, UKL} or embedded systems~\cite{PRACTICAL_OS_TAILORING, ML_CONF_SIZE}, and also affect aspects such as security~\cite{ISOLSEC, ISOLSEC2}.
To optimize for memory consumption we configure \sys to build images of RISC-V Linux, an ISA that is highly popular in embedded systems.
The memory consumption is measured by booting the images in a QEMU emulated setup (although emulation affects performance, it does not impact memory consumption).
In line with past studies on that topic~\cite{LUPINE, ISOLSEC, ISOLSEC2, PRACTICAL_OS_TAILORING, ML_CONF_SIZE}, in this experiment we configure \sys to favor varying compile-time options (over run-time options as was the case in \S\ref{sec:eval-perf}).
We launch the exploration process with \sys and random search, for a time budget of 3 hours.

\begin{figure}
    \centering
    \includegraphics[width=.45\textwidth]{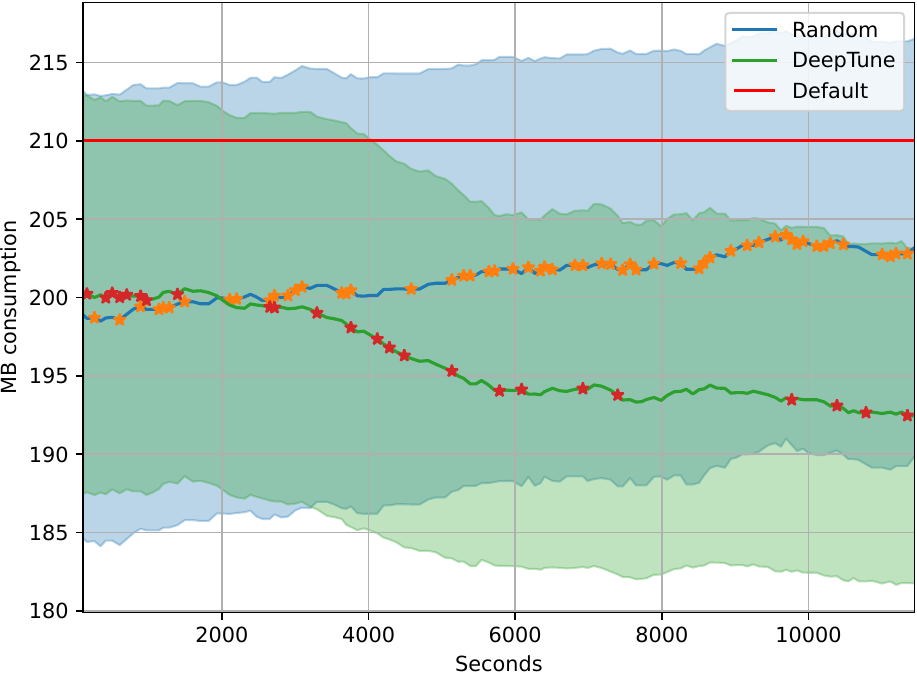}
    \caption{Memory footprint of Linux images built from configurations found during a 3-hour search session by \sys and random search. Results of 5 runs, smoothed for readability. Crashes are indicated with stars.}
    \label{fig:memory-footprint}
\end{figure}

The evolution of the memory footprint of the configurations found is presented on \Cref{fig:memory-footprint}.
The default configuration has a 210 MB memory footprint.
After 3 hours, \sys finds a configuration having a memory footprint of 192 MB (smoothed), which is an 8.5\% reduction with respect to the default one.
At the end of the exploration process, the memory footprint of the configuration found by random search is 203 MB (smoothed), i.e. a reduction of 5.5\%.
Further, one can notice that the number of failing configurations is much smaller with \sys than with random search.
This is due to the system's capacity to predict crashes, which is particularly efficient for that experiment: only four crashes happen in the last 100 minutes.

\begin{figure}
    \centering
    \includegraphics[width=.5\textwidth]{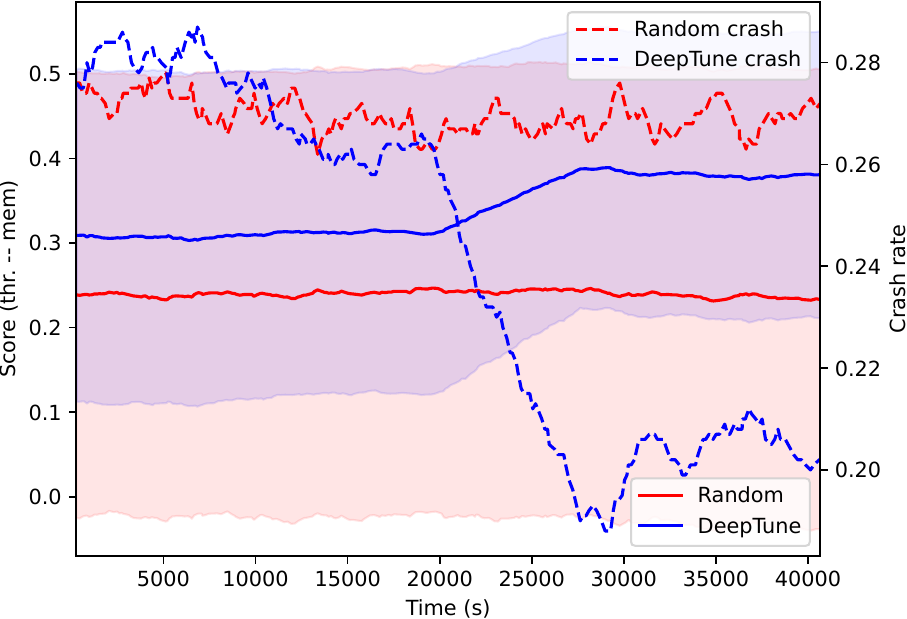}
    \caption{Evolution of a learning policy that co-optimizes throughput and memory consumption on top of Cozart. Higher is better. Application under test is Nginx. Smoothed for readability. The crash rate indicated on a separate axis.}
    \label{fig:cozart}
\end{figure}

\paragraph{Synergy with Compile-Time Optimizers.}
Using the methods discussed earlier, \sys can also examine compile-time parameters. However, this process can be inefficient due to how the Linux kernel manages build dependencies and how \sys explores configurations. Initially, \sys generates nearly random configurations, which often requires complete rebuilds of the Linux kernel. Even a single incorrect parameter can cause the build to fail. To address this issue, we added an initial optimization step using Cozart~\cite{COZART}, which uses dynamic analysis to significantly reduce the number of unused components in the Linux kernel. This reduction results in a much smaller configuration space, allowing \sys to focus on the unexplored runtime parameters of Linux while also providing a performance boost. We observed a 31\% increase in throughput compared to the baseline, along with a slight decrease in memory usage, similar to what was reported in the Cozart evaluation.

Next, we applied \sys optimizations on top of the Cozart baseline. We define an optimization score $s$ as follows:
\begin{equation}
s = \text{mXNorm}(t) - \text{mXNorm}(m),
\end{equation}
Here, $t$ represents throughput, $m$ represents memory consumption, and \text{mXNorm}(x) is the min-max normalization function, which brings throughput and memory numbers to a common scale.
A higher score thus indicates better throughput and less memory usage.

Figure~\ref{fig:cozart} shows that \sys can effectively learn and create a policy that performs better than random search when run on top of a Cozart baseline.
Similarly to Figure~\ref{fig:unikraft-nginx}, this figure shows the exploration vs. exploitation strategies of \algo.
After about 300 minutes, \algo finds a set of parameters that allows it to increase the throughput-memory score.
It then exploits it for about 100 minutes.
This manifests in a significantly lower crash rate since \algo focuses on a more stable part of the space.
\algo switches back to exploration after this phase.
This results in a higher crash rate as it explores more unknown configurations.

% Table \ref{tab:wayfinder-cozart} lists the top five scores found during the joint exploration of throughput and memory. The first permutation, while having significantly lower throughput, also uses less memory, resulting in the highest score. In contrast, the second permutation achieves much higher throughput than the first and shows a slight improvement over the Cozart baseline, while also reducing memory usage by more than 4 MB.
Table \ref{tab:wayfinder-cozart} presents the top five scores obtained during the joint exploration of throughput and memory usage.
The first and second permutations, despite having significantly lower throughput, utilize less memory, leading to the highest overall scores.
Conversely, the third permutation achieves much higher throughput with only a slight increase in memory usage.
Compared to the Cozart baseline, these permutations consistently deliver higher throughput while maintaining lower memory consumption.

In conclusion, we demonstrate that by combining the Cozart compilation optimizations with the runtime optimizations from \sys, the operating system can be adjusted to improve performance across various metrics.

% \begin{table}
%     \center

%     \caption{Top 5 best results found for the Throughput-Memory experiment (Figure \ref{fig:cozart}) on top of Cozart.}

%     \footnotesize
%     \begin{tabular}{|c | c| c | c|}
%     \hline
%      {\bf Rank} & Score &  Memory (mB) & Throughput (req./s)  \\
%       \hline
%       1  & 0.84 & 323.52  & 40868 \\
%       \hline
%       2  & 0.82 & 327.59  & 48623 \\
%       \hline
%       3 & 0.81   & 327.53  & 48268 \\
%       \hline
%       4    & 0.79  & 327.48   & 47073 \\
%       \hline
%       5    & 0.78  & 327.54   & 47002 \\
%       \hline \hline
%       Cozart  & --  & 331.77   & 46855 \\
%       \hline
%     \end{tabular}

%     \label{tab:wayfinder-cozart}
% \end{table}

\begin{table}
    \center

    \caption{
Top-5 best results found for the Throughput-Memory experiment (Figure \ref{fig:cozart}) on top of Cozart. Note that these numbers are obtained with the same setup as the Cozart paper's~\cite{COZART}, and thus cannot be compared with Table~\ref{tab:default-conf} (different kernel versions, four CPU cores instead of one).}

    \footnotesize
    \begin{tabular}{|c | c| c | c|}
    \hline
     {\bf Rank} & Score &  Memory (mB) & Throughput (req./s)  \\
      \hline
      1  & 0.84 & 327.72  & 47002 \\
      \hline
      2  & 0.82 & 328.57  & 47215 \\
      \hline
      3 & 0.81   & 329.67  & 49375 \\
      \hline
      4    & 0.79  & 329.75   & 48606 \\
      \hline
      5    & 0.78  & 330.46   & 49375 \\
      \hline \hline
      Cozart  & --  & 331.77   & 46855 \\
      \hline
    \end{tabular}

    \label{tab:wayfinder-cozart}
\end{table}

%% file: 05-related-works.tex
\section{Related Works}
\label{sec:related-works}

\paragraph{Configuration-based OS Specialization.}
Several works specialize Linux through its configuration to lower resource usage~\cite{PRACTICAL_OS_TAILORING, ML_CONF_SIZE, LUPINE, UKL, XCONTAINER, LOHMANN} or attack surface~\cite{ISOLSEC, ISOLSEC2}.
Some approaches are manual~\cite{XCONTAINER, LUPINE, UKL} and cannot scale to many applications, while others are automated.
Undertaker~\cite{ISOLSEC, ISOLSEC2} uses dynamic analysis to trace kernel code executed by a workload, and correlates that code to compile-time options to compile out as much unneeded code as possible.
Kernel tailoring~\cite{PRACTICAL_OS_TAILORING} builds on top of Undertaker.
Noting that a large part of the configurations explored by the tool fail at runtime, the work proposes a search method that tries to derive valid configurations from configurations known to execute successfully.
These approaches target attack surface/memory footprint reduction, so they aim to determine, for a given workload, which compile-time configuration parameters can be switched off, and which must be on.
These approaches cannot be directly applied to our problem due to our focus on performance (in addition to other metrics such as resource usage).
Beyond turning only compile-time features on and off, \sys considers the full set of kernel compile-time, boot-time, and run-time options, many of them taking arbitrary values, with undocumented ranges of validity.
Cozart~\cite{COZART} also builds on top of Undertaker.
With a similar approach that uses dynamic analysis to determine unused configuration options, Cozart reduces the footprint of the Linux kernel, with the added bonus of decreasing memory usage and increasing performance.
These performance gains are not the main focus of Cozart though, but they do offer a better baseline for \sys to start from.
As mentioned previously, \sys and Cozart do not have much in common, but they synergize very well when it comes to optimizing the Linux kernel both through compile-time options and through run-time ones. 
AutOS~\cite{AUTOS} specializes OS kernels (including Linux) for AIoT application performance, by expressing their configuration as a tree and passing it to a large language model driving the exploration with the help of prompts describing the environment and refining the search.
This work is limited to compile-time options and requires expertise in order to write prompts.

\paragraph{Configuration-based Application Optimization.}
Unicorn~\cite{UNICORN} uses causal inference to automatically specialize configurations for embedded software.
KML~\cite{KML1, KML2} uses machine learning techniques to optimize the configuration of storage subsystems.
Some past works have explored the performance impact of software configuration parameters using static and/or dynamic analysis~\cite{PERF_EXPLO_STATIC, PERF_EXPLO_DYN1, PERF_EXPLO_DYN2}.
Others~\cite{CARVER} focus on identifying the configuration parameters most relevant to performance for a certain workload, in this case in a storage subsystem.
All these systems require expert knowledge to pre-identify sets of potentially important parameters, and target configuration spaces that are much smaller than \sys's.
In particular, we have shown (Fig.~\ref{fig:Scalability}) that Unicorn does not scale to such large configuration spaces.

In serverless computing, AWS Lambda Power Tuning~\cite{AWSPT1, AWSPT2} is an automated platform to optimize the memory allocated to functions.
Sizeless~\cite{SIZELESS} proposes a regression model to estimate function performance.
COSE~\cite{COSE} uses Bayesian optimization to find specialized configurations for functions.
In databases, several works such as Ottertune~\cite{OTTERTUNE1} also automatically optimize for performance.
Such application-specific approaches present a configuration space that is orders of magnitude smaller than that of modern OSes.
Hence, they rely on techniques such as Bayesian optimization that do not fit our problem.

Configuration-based optimization is used in other contexts including hardware design (accelerators tuning~\cite{ACCELERATOR1, ACCELERATOR2}, FPGA floor planning~\cite{FPGA_FLOORPLAN}), neural architecture search~\cite{NEURAL_ARCH}, and distributed computation placement~\cite{DISTRIBUTED_PLACEMENT}.
These methods are tied to the application domain they target (e.g. they operate on a specific representation of the search space~\cite{NEURAL_ARCH}) and are not directly applicable to OS specialization.

\paragraph{Other Optimization Works in the ML Literature.}
Parameter search/optimization has attracted significant attention in recent years~\cite{hase2018, hase2018chimera, pollice2021data, friederich2021machine, UNICORN, SMAC3, akiba2019optuna, SnoekRSKSSPPA15, feurer2015efficient}.
Some approaches rely on Bayesian optimization.
SMAC~\cite{SMAC3}, for example, enables to easily apply Bayesian optimization to generic optimization problems.
SMAC also supports random forests, which improve on scalability and performance prediction compared to a Gaussian process, but offer poor estimations of predicted posterior uncertainty (needed for exploration vs. exploitation decisions).
Snoek et al.~\cite{SnoekRSKSSPPA15} contribute Bayesian Neural Networks (BNN) to replace Gaussian processes.
BNNs are general approximators that can predict the posterior uncertainty, but their fitting and generalization capabilities are typically lower than conventional NNs.
BNNs are also expensive to evaluate, since they require a Monte Carlo sampling strategy that requires to run them multiple times.
Hase et al.~\cite{hase2018} successfully apply this idea in chemistry, and extend the idea to multi-objective optimization~\cite{hase2018chimera}.
Gal et al.~\cite{gal2016dropout} sample from dropout as an approximation of BNNs.
Finally, Kendall et al.~\cite{KendallG17} study several kinds of uncertainties required for ML problems, and propose an uncertainty estimation mechanism for regression problems.
Due to the nature of the conventionally used activation functions, they typically cannot provide a good uncertainty of outlier samples.
\algo proposes a new NN design that enables accurate prediction of system performance, failure, and accuracy in a single pass with high precision.
We also propose a new scoring function that creates a balanced permutation policy picking. 

%% file: 06-conclusion.tex
\section{Conclusion}
\label{sec:conclusion}

We proposed \sys, a fully automated approach at OS specialization through configuration.
\sys ships with a benchmarking platform, and \algo, a neural-network-based optimization algorithm which predicts the validity and performance of configurations.
\sys does not require expert knowledge, and scales to the vast configuration space of modern OSes such as Linux.
After running \sys to optimize for one application, the trained model can be reused to speed up the specialization for other similar applications.
\sys can identify specialized configurations for various metrics, with up to 24\% application performance increase and 8.5\% memory usage decrease vs. default configurations, converging faster on good solutions than competing approaches using random search and Bayesian optimization.

%% file: 07-acknowledgements.tex
\section*{Acknowledgments}

We thank the anonymous reviewers for their comments and insights.
Special thanks go to Margo Seltzer for her precious comments that significantly improved the paper's quality.
This work was funded in part by Innovate UK grant 10164504 (MicrOS), and the UK Engineering and Physical Sciences Research Council grants EP/V012134/1 (UniFaaS), EP/V000225/1 (SCorCH), and EP/X015610/1 (FlexCap).
We acknowledge the support of the Natural Sciences and Engineering Research Council of Canada (NSERC).
Nous remercions le Conseil de recherches en sciences naturelles et en génie du Canada (CRSNG) de son soutien.

%% file: A-artifact-evaluation.tex
%%%%%%%%%%%%%%%%%%%%%%%%%%%%%%%%%%%%%%%%%%%%%%%%%%%%
% Artifact Appendix Template for EuroSys'26 AE
%
% this document has a maximum length of 2 pages.
%%%%%%%%%%%%%%%%%%%%%%%%%%%%%%%%%%%%%%%%%%%%%%%%%%%%

\appendix
\section{Artifact Appendix}

%%%%%%%%%%%%%%%%%%%%%%%%%%%%%%%%%%%%%%%%%%%%%%%%%%%%%%%%%%%%%%%%%%%%%
\subsection{Abstract}
{\em This artifact contains the source code of \sys, the operating system specialization framework.
It also contains scripts to reproduce the results found in the paper, and/or the dataset used for generating the figures/tables in the paper.
The goal is to allow the reader to reproduce our experiments, and to give insight into the capabilities and usage of \sys.}

%%%%%%%%%%%%%%%%%%%%%%%%%%%%%%%%%%%%%%%%%%%%%%%%%%%%%%%%%%%%%%%%%%%%%
\subsection{Description \& Requirements}

\subsubsection{How to access}
The latest version of the artifact can be found on GitHub\footnote{\href{https://github.com/unikraft/wayfinder-eurosys26-ae}{\textsf{https://github.com/unikraft/wayfinder-eurosys26-ae}}}.
Alternatively, individual releases can be downloaded from our Zenodo archive\footnote{\href{https://doi.org/10.5281/zenodo.18598079}{\textsf{https://doi.org/10.5281/zenodo.18598079}}}.
Note that the artifact evaluation (AE) GitHub repository only contains part of the artifact, namely scripts to reproduce this paper's experiments.
The \sys framework code, tools, and other examples are available in the \sys\footnote{\href{https://github.com/unikraft/wayfinder}{\textsf{https://github.com/unikraft/wayfinder}} and\\\href{https://doi.org/10.5281/zenodo.18592520}{\textsf{https://doi.org/10.5281/zenodo.18592520}}} repository.

In order to precisely reproduce this paper's measurements, we gave EuroSys'26 AEC reviewers access to our server, an Intel(R) Xeon(R) CPU E5-2690 v3 @ 2.60GHz with 315 GB RAM, Ubuntu 24.04, and Linux version 6.8.0-53-generic. Access to this particular setup is not required to run the artifact. Hardware and software dependencies are detailed below.

\subsubsection{Hardware dependencies}
An Intel(R) Xeon(R) CPU E5-2690 v3 @ 2.60 GHz with at least 64 GB RAM, typically any processor which has a dual-socket mount.
The processor must have at least 8 cores.
64 GB of RAM are necessary to run the experiments corresponding to Figure \ref{fig:Scalability}, as the experiment tests RAM usage.
Note that this amount of cores/RAM is required to reproduce this paper's results, not to run \sys.
At least 100 GB of disk space is needed.

\subsubsection{Software dependencies}
\label{sec:software-deps}
This artifact has been tested with Ubuntu 24.04 (Noble Numbat), with Linux kernel version 6.8.0-53-generic (KVM enabled), Docker version 28.1.1 (or any recent version), QEMU 5.2, Libvirt 7.4.
Other dependencies use containers which are already versioned by \sys and the artifact scripts.

\subsubsection{Benchmarks}
All data sets are included in the artifact, either generated automatically by scripts, or pre-genera\-ted.

%%%%%%%%%%%%%%%%%%%%%%%%%%%%%%%%%%%%%%%%%%%%%%%%%%%%%%%%%%%%%%%%%%%%%
\subsection{Set-up}
\label{sec:set-up}

Before running any experiment, prepare your host with the recommendations detailed above in \ref{sec:software-deps}.
Once the system is set up, clone the AE repository:

\begin{console}
$ git clone \
https://github.com/unikraft/wayfinder-eurosys26-ae
\end{console}

Then follow the \emph{README.md} step-by-step instructions to download, compile, and configure \sys from its repository:

\begin{console}
$ git clone https://github.com/unikraft/wayfinder
\end{console}

Finally, after everything is configured, start \sys and its helper containers:

\begin{console}
$ docker-compose up -d registry influxdb postgres redis
$ sudo ./scripts/wayfinder.sh
\end{console}

%%%%%%%%%%%%%%%%%%%%%%%%%%%%%%%%%%%%%%%%%%%%%%%%%%%%%%%%%%%%%%%%%%%%%
\subsection[Evaluation workflow]{Evaluation workflow\footnote{Submission, reviewing and badging methodology followed for the evaluation of this artifact can be found at \href{https://sysartifacts.github.io/eurosys2026/}{\textsf{https://sysartifacts.github.io/eurosys2026/}}.}}

Experiments should be run sequentially.
Every experiment uses a different builder image to generate kernels to test.
These images need to be built before running the experiment, for example:

\begin{console}
$ docker build -t localhost:5000/linux-nginx:latest -f Dockerfile.jessie .
$ docker push localhost:5000/linux-nginx:latest
\end{console}

The preparation of each experiment is a mandatory step that consists in building the kernels that will be tested by the platform.
Each kernel build operation only takes minutes and should be done while no other experiment is running on the system.
Once one of the experiments has been prepared, it can be run with a syntax similar to this one:

\begin{console}
$ ~/wayfinder/dist/wfctl create ./job.yaml
$ ~/wayfinder/dist/wfctl start --isol-level full --isol-split both -l 1 -s random "$ID"
\end{console}

Running all experiments takes on average days on our setup.
This is because we collect thousands of permutations to showcase the evolution.
For the experiments that interact with \sys, consider decreasing the number of permutations and repetitions.
For the rest, we provide pre-generated data-sets that can be used to quickly generate the figure.

\subsubsection{Major Claims}

\begin{itemize}
    \item \textit{(C1): \sys fully automatically identifies specialized configurations with up to 24\% application performance improvement and 8.5\% memory usage reduction compared to default configurations. This is proven by the experiments described in Sections \ref{sec:eval-perf} and \ref{sec:eval-misc} whose results are illustrated in \Cref{fig:perf-general} and \Cref{fig:memory-footprint}.}
    \item \textit{(C2): We highlight the benefits of our neural network, reaching good solutions faster than competing approaches (random and Bayesian), and successfully transferring knowledge between related applications. This is proven by the experiments described in Section \ref{sec:eval-misc} whose results are illustrated in \Cref{fig:unikraft-nginx} and \Cref{fig:cozart}.}
\end{itemize}

\subsubsection{Experiments}

Experiments presented and structured in the artifact repository mentioned in \Cref{sec:set-up} each have a \emph{README.md} file and are structured suggestively in directories of the formats \emph{figure-*} and \emph{table-*}.
These files briefly describe the experiments and the relevant preparation and execution steps: it is strongly recommended to go through them at least once.
Experiments which do not have a time estimate take minutes.
The total time to run the experiments in the repository should be around 10-12 hours, but can take up to 32 hours if repetitions are done.
Out of this, in both cases, only 2-3 hours represent manual work.
The \sys framework takes care to run jobs in the background.

Reproducing experiments on the same machine should produce results similar as in the paper.
On other machines, we expect different absolute numbers, but similar trends and ordering.
Because of the non-deterministic nature of the exploration process, repeated measurements are subject to some variation, but the general trends and averages of multiple executions should be consistent with what is presented in the paper.
Figures \ref{fig:linux-plot}, \ref{fig:linux_throughput} and Table \ref{tab:linux-parameter-types}, are reproducible manually.
Due to the very long exploration time (days) required to produce the data behind them, Figures \ref{fig:tl-matrix}, \ref{fig:perf-general}, \ref{fig:Scalability}, \ref{fig:exploration-loop}, \ref{fig:unikraft-nginx}, \ref{fig:memory-footprint}, \ref{fig:cozart} are reproducible from existing datasets. 
Finally, the measurements in the data use the following metrics: \emph{requests/s, ms/op, Mop/s, seconds, MB}. 

\subsection{Notes on Reusability}
\label{sec:reuse}
Reviewers may use the examples in the \sys repository to create their own job based on a custom application and its accompanying benchmarking tool.
Instructions to build the \sys developer Docker image, run generic jobs, and build custom images are available in the \emph{README.md} file at the root of our AE repository\footnote{\href{https://github.com/unikraft/wayfinder-eurosys26-ae/blob/main/README.md}{\textsf{https://github.com/unikraft/wayfinder-eurosys26-ae/blob/main/README.md}}}.

%%%%%%%%%%%%%%%%%%%%%%%%%%%%%%%%%%%%%%%%%%%%%%%%%%%%%%%%%%%%%%%%%%%%%
\subsection{General Notes}
\label{sec:gnotes}
All experiments have a \emph{README.md} file detailing the manual steps that are needed to obtain the corresponding figures.
The folders for Figure \ref{fig:linux-plot}, Table \ref{tab:linux-parameter-types}, and Figure \ref{fig:linux_throughput} contain the step-by-step guide to generate them from scratch.
The other experiments contain the data sets obtained and scripts to generate the Figures/Tables in question.
Most experiments took at least one day to run each, so it is recommended to inspect and adapt the scripts provided to lower the number of iterations.
We strongly recommend carefully reading the instructions in each experiment before starting to reproduce experiments.

%%%%%%%%%%%%%%%%%%%%%%%%%%%%%%%%%%%%%%%%%%%%%%%%%%%%%%%%%%%%%%%%%%%%%